\documentclass[twocolumn,preprintnumbers,showpacs,amsmath,amssymb]{revtex4}
\usepackage{graphicx}
\usepackage{amssymb}

\begin{document}
\title{Secure quantum bit commitment against empty promises. II. The density
matrix}
\author{Guang Ping He}
\email{hegp@mail.sysu.edu.cn}
\affiliation{School of Physics and Engineering, Sun Yat-sen University, Guangzhou 510275,
China}

\begin{abstract}
We further study the security of the quantum bit commitment (QBC) protocol
we previously proposed [\textit{Phys. Rev.} A \textbf{74}, 022332 (2006).],
by analyzing the reduced density matrix $\rho _{b}^{B}$ which describes the
quantum state at Bob's side corresponding to Alice's committed bit $b$. It
is shown that Alice will find $\rho _{0}^{B}\perp \rho _{1}^{B}$ while the
protocol remains concealing to Bob. On the contrary, the existing no-go
theorem of unconditionally secure QBC is based on the condition $\rho
_{0}^{B}\simeq \rho _{1}^{B}$. Thus the specific cheating strategy proposed
in the no-go theorem does not necessarily applies to our protocol.
\end{abstract}

\maketitle

\newpage


\section{Introduction}

Quantum bit commitment (QBC) is a two-party cryptography including two
phases. In the commit phase, Alice (the sender of the commitment) decides
the value of the bit $b$ ($b=0$ or $1$) that she wants to commit, and sends
Bob (the receiver of the commitment) a piece of evidence, e.g., some quantum
states. Later, in the unveil phase, Alice announces the value of $b$, and
Bob checks it with the evidence. An unconditionally secure QBC protocol
needs to be both binding (i.e., Alice cannot change the value of $b$ after
the commit phase) and concealing (Bob cannot know $b$ before the unveil
phase) without relying on any computational assumption.

It is widely accepted that unconditionally secure QBC is impossible \cite%
{qi74}-\cite{qbc31}, despite of some attempts towards secure ones (a
detailed list and brief history can be found in the introduction of \cite%
{HeJPA}). This result, known as the Mayers-Lo-Chau (MLC) no-go theorem, was
considered as putting a serious drawback on quantum cryptography.

Nevertheless, we must note that the correctness of the \textit{conclusion} a
theorem should not be confused with that of its \textit{proof}. While a
correct proof will surely lead to a correct conclusion, there could also be
cases where someone may draw a correct conclusion despite that the existing
proof is not sufficiently general. In quantum cryptography, though there are
brilliant proofs (e.g., \cite{qi70}) for the security of quantum key
distribution, for other cryptographic tasks it could be hard to find a
general proof showing that a protocol is unconditionally secure, since there
could potentially exist numerous cheating strategies. Similarly, it is also
hard to find a real general proof showing that a cryptographic task can
never be accomplished securely (unless the definition of the task contains
self-inconsistent goals), because the protocols potentially
existed could also be numerous, some of which may even beyond our current
imagination. As for QBC, it is important to notice that all the existing
no-go proofs \cite{qi74}-\cite{qbc31} are actually based on a specific
cheating strategy of Alice, as it will be summarized below. No matter
unconditionally secure QBC is possible or not, we could question whether
this specific cheating strategy can be evaded. If there is a protocol which
is secure against the specific cheating strategy in the no-go proofs while
insecure against other cheating strategies, then it reveals that the
existing \textit{proofs} of the MLC no-go theorem should not be considered
sufficiently general, despite that the \textit{conclusion} of the theorem
may remain valid.

In our previous work \cite{He}, we proposed a QBC protocol and proved that
it is secure against some known attacks, while an attack strategy that can
break our protocol successfully has yet to be found. Thus the exact boundary
of the security of the protocol remains unclear. In this paper, we will
further show that the density matrix in the protocol displays a
distinct feature comparing with that of the QBC model studied in
existing no-go proofs \cite{qi74}-\cite{qbc31}. This makes it possible for
our protocol to evade at least the specific cheating strategy that
led to these proofs.

In the next section, we will briefly review the existing no-go proofs of
QBC, and pinpoint out that the cheating strategies in all these proofs have
the same requirement on the density matrix. Our previous QBC protocol \cite%
{He} will be illustrated in section III. Then in section IV, we will analyze
the density matrix in this protocol, and show that they does not satisfy a
requirement on which the no-go proofs hold. In section V, we will elaborate
why security can maintain in the absence of this requirement.

\section{The density matrix in the no-go proofs}

Although there are many no-go proofs \cite{qi74}-\cite{qbc31}, they all
have the following common features.

(1) The reduced model. According to the no-go proofs, any QBC protocol can
be reduced to the following model. Alice and Bob together own a quantum
state in a given Hilbert space. Each of them performs unitary
transformations on the state in turns. All measurements are performed at the
very end.

(2)\ The coding method. The quantum state corresponding to the committed bit
$b$ has the form%
\begin{equation}
\left\vert \psi _{b}\right\rangle =\sum\limits_{j}\lambda
_{j}^{(b)}\left\vert e_{j}^{(b)}\right\rangle _{A}\otimes \left\vert
f_{j}^{(b)}\right\rangle _{B}\text{ ,}  \label{eqMLC}
\end{equation}%
and it is known to both Alice and Bob. Here the systems $A$ and $B$ are
owned by Alice and Bob respectively.

(3) The concealing condition. To ensure that Bob's information on the
committed bit is trivial before the unveil phase, any QBC protocol secure
against Bob should satisfy%
\begin{equation}
\rho _{0}^{B}\simeq \rho _{1}^{B},  \label{concealing}
\end{equation}%
where $\rho _{b}^{B}=Tr_{A}\left\vert \psi _{b}\right\rangle \left\langle
\psi _{b}\right\vert $ is the reduced density matrix of the state at Bob's
side corresponding to Alice's committed bit $b$.

(4) The cheating strategy. Once Eq. (\ref{concealing}) is satisfied,
according to the Hughston-Jozsa-Wootters (HJW) theorem (which also appeared
in many different names in literature, e.g., the Uhlmann theorem, etc.) \cite%
{steering,steering 2,qbc8,qi73}, there exists a local unitary transformation
for Alice to map $\{\left\vert e_{j}^{(0)}\right\rangle _{A}\}$ into $%
\{\left\vert e_{j}^{(1)}\right\rangle _{A}\}$ successfully with a high
probability. Thus a dishonest Alice can unveil the state as either $%
\left\vert \psi _{0}\right\rangle $ or $\left\vert \psi _{1}\right\rangle $\
at her will with a high probability to escape Bob's detection. For this
reason, a concealing QBC protocol cannot be binding.

The most important point for our discussion here is feature (3). We would
like to emphasize again that it appears in all existing no-go proofs. Note
that in some references (e.g. \cite{qi147,qi714,qbc31,qi323}), this feature
was expressed using the trace distance or the fidelity instead of the reduced
density matrices, while the meaning remains the same. On the other hand, it
will be shown below that the density matrix in our previous QBC protocol
\cite{He} displays an intriguing feature. Though the protocol remains
concealing against Bob, at Alice's point of view there will be $\rho
_{0}^{B}\perp \rho _{1}^{B}$ (i.e., they are orthogonal) instead of $\rho
_{0}^{B}\simeq \rho _{1}^{B}$. As Eq. (\ref{concealing}) is necessary for
constructing Alice's cheating transformation in the above feature (4), our
protocol is thus immune to this specific cheating strategy.

\section{Our protocol}

\subsection{The rigorous description}

In Ref. \cite{He}, we proposed the following QBC protocol.

\bigskip

The \textit{commit} protocol: [$commit(b)$]

(C1) Alice and Bob first agree on a security parameter $s$, then $%
DO_{i=1}^{s}$ Alice picks $\theta _{i}\in (0,\pi /2)$ ($\theta _{i}$ needs
not to be different for each $i$. For example, Alice can fix $\theta
_{i}=\pi /4$\ throughout the whole protocol) and randomly picks $q_{i}\in
\{0,1\}$, and prepares an entangled state
\begin{equation}
\left\vert \psi _{i}\right\rangle =\left\vert \alpha _{i}\otimes \beta
_{i}\right\rangle =\cos \theta _{i}\left\vert x\right\rangle _{\alpha
}\otimes \left\vert 0,q_{i}\right\rangle _{\beta }+\sin \theta
_{i}\left\vert y\right\rangle _{\alpha }\otimes \left\vert
1,q_{i}\right\rangle _{\beta }.  \label{eqA1}
\end{equation}%
Then she sends the quantum register $\beta _{i}$ to Bob and stores $\alpha
_{i}$. Here $\left\vert x\right\rangle _{\alpha }$ and $\left\vert
y\right\rangle _{\alpha }$ are two orthogonal states of the quantum register
$\alpha _{i}$, while we use $\left\vert p_{i},q_{i}\right\rangle _{\beta }$
to denote the state of $\beta _{i}$, with $p_{i}$ denoting the basis and $%
q_{i}$\ labelling the different states in the same basis. The state $%
\left\vert 0,0\right\rangle $ and $\left\vert 0,1\right\rangle $ are
orthogonal to each other, and $\left\vert 1,0\right\rangle \equiv
(\left\vert 0,0\right\rangle +\left\vert 0,1\right\rangle )/\sqrt{2}$, $%
\left\vert 1,1\right\rangle \equiv (\left\vert 0,0\right\rangle -\left\vert
0,1\right\rangle )/\sqrt{2}$;

(C2) Bob chooses a number $s^{\prime }$ ($0\leq s^{\prime }<s$) and
randomly divides $S\equiv \{1,...,s\}$ into two subsets $S^{\prime }$ and $%
S^{\prime \prime }$ such that $\left| S^{\prime }\right| =s^{\prime }$, $%
S^{\prime \prime }=S-S^{\prime }$. Then for $\forall i\in S^{\prime }$ Bob
stores $\beta _{i}$\ unmeasured. And for $\forall i\in S^{\prime \prime }$
Bob randomly picks a basis $p_{i}^{\prime }\in \{0,1\}$ and measures $\beta
_{i}$. The outcome is denoted as $\left| p_{i}^{\prime },q_{i}^{\prime
}\right\rangle _{\beta }$;

(C3) Bob chooses $f_{a}$, $f_{b}$, $f_{c}$ ($f_{a}+f_{c}<1/2$ and $%
f_{b}>f_{c}$) and announces to Alice the \textquotedblleft
fake\textquotedblright\ result $\{\left\vert p_{i}^{\prime \prime
},q_{i}^{\prime \prime }\right\rangle _{\beta }|i\in S\}$ such that $%
f_{a}=(\left\vert L_{a}\right\vert +s^{\prime }/4)/s$, $f_{b}=(\left\vert
L_{b}\right\vert +s^{\prime }/4)/s$ and $f_{c}=(\left\vert L_{c}\right\vert
+s^{\prime }/4)/s$, where $L_{a}=\{i\in S^{\prime \prime }|$ $\left\vert
p_{i}^{\prime \prime },q_{i}^{\prime \prime }\right\rangle _{\beta }=$ $%
\left\vert p_{i}^{\prime },\urcorner q_{i}^{\prime }\right\rangle _{\beta
}\} $, $L_{b}=\{i\in S^{\prime \prime }|$ $\left\vert p_{i}^{\prime \prime
},q_{i}^{\prime \prime }\right\rangle _{\beta }=$ $\left\vert \urcorner
p_{i}^{\prime },q_{i}^{\prime }\right\rangle _{\beta }\}$, and $L_{c}=\{i\in
S^{\prime \prime }|$ $\left\vert p_{i}^{\prime \prime },q_{i}^{\prime \prime
}\right\rangle _{\beta }=$ $\left\vert \urcorner p_{i}^{\prime },\urcorner
q_{i}^{\prime }\right\rangle _{\beta }\}$;

(C4) Alice divides $S$ into two subsets: $M=\{i\in S|q_{i}^{\prime \prime
}=\urcorner q_{i}\}$ and $U=\{i\in S|q_{i}^{\prime \prime }=q_{i}\}$. For $%
\forall i\in M$, she measures $\alpha _{i}$ in the basis $(\left\vert
x\right\rangle _{\alpha },\left\vert y\right\rangle _{\alpha })$. She sets $%
p_{i}=0$ if she finds $\left\vert x\right\rangle _{\alpha }$ or $p_{i}=1$ if
she finds $\left\vert y\right\rangle _{\alpha }$. Then she sets $L=\{i\in
M|p_{i}=p_{i}^{\prime \prime }\}$ and announces it to Bob; (Since it could be shown
that $\left\vert M\right\vert \simeq \lbrack 1/4+(f_{a}+f_{c})/2]s$, by checking
whether $\left\vert M\right\vert <s/2$ Alice can test whether Bob has indeed
chosen $f_{a}+f_{c}<1/2$. Also, since $\left\vert L\right\vert \simeq
(f_{a}/2+f_{b}/4+f_{c}/4)s$, we have $\left\vert M\right\vert -\left\vert
L\right\vert \simeq \lbrack 1/4-(f_{b}-f_{c})/4]s$. Thus by checking whether
$\left\vert M\right\vert -\left\vert L\right\vert <s/4$\ Alice can test
whether Bob has indeed chosen $f_{b}>f_{c}$.)

(C5) Bob sets $L_{s^{\prime }}=L\cap S^{\prime }$. Then he measures $\beta
_{i}$\ ($\forall i\in L_{s^{\prime }}$) in the basis $p_{i}^{\prime
}=p_{i}^{\prime \prime }$ and denotes the outcome as $\left\vert
p_{i}^{\prime },q_{i}^{\prime }\right\rangle _{\beta }$. He agrees to
continue only if $\{i\in L_{s^{\prime }}|\left\vert p_{i}^{\prime
},q_{i}^{\prime }\right\rangle _{\beta }=\left\vert p_{i}^{\prime \prime
},q_{i}^{\prime \prime }\right\rangle _{\beta }\}=\phi $, $L\subset
L_{a}\cup L_{b}\cup L_{c}\cup S^{\prime }$ and $\left\vert L\right\vert
\simeq (f_{a}/2+f_{b}/4+f_{c}/4)s$;

(C6) Alice sets $c_{i}^{0}=0$ if $i\in U$ or $c_{i}^{0}=1$ if $i\in M-L$.
Thus she obtains a binary string $c^{0}=(c_{1}^{0}c_{2}^{0}...c_{n}^{0})$ ($%
n\equiv \left| S-L\right| $);

(C7) Alice and Bob complete the commitment with the codeword method similar
to that of the BCJL protocol \cite{qi43} by using $c^{0}$ to encode the
codeword ($c^{0}$ itself is not announced to Bob). That is:

\qquad (C7.1) Bob chooses a
binary linear $(n,k,d)$-code $C$ and announces it to Alice, where the ratios
$d/n$ and $k/n$ are agreed on by both Alice and Bob;

\qquad (C7.2) Alice chooses a nonzero random $n$-bit string $%
r=(r_{1}r_{2}...r_{n})\in \{0,1\}^{n}$ and announces it to Bob;

\qquad (C7.3) Now Alice has in mind the value of the bit $b$ that she wants
to commit. Then she chooses a random $n$-bit codeword $%
c=(c_{1}c_{2}...c_{n}) $ from $C$ such that $c\odot r=b$ (Here $c\odot
r\equiv \bigoplus\limits_{i=1}^{n}c_{i}\wedge r_{i}$);

\qquad (C7.4) Alice announces to Bob $c^{\prime }=c\oplus c^{0}$.

\bigskip

The \textit{unveil} protocol: [$unveil(b,c,c^{0},\left\vert \psi
_{i}\right\rangle )$]

(U1) Alice announces $b$, $c$, $c^{0}$, $\{q_{i},\theta _{i}|$ $i\in S\}$
and $\{p_{i}|i\in M\}$\ to Bob;

(U2) Alice sends the quantum registers $\{\alpha _{i}|i\in U\}$ to Bob;

(U3) Bob finishes the measurement on $\{\alpha _{i}|i\in U\}$ and $\{\beta
_{i}|i\in S^{\prime }\}$ to check Alice's announcement;

(U4) Bob checks $\left\vert M\right\vert \simeq \lbrack
1/4+(f_{a}+f_{c})/2]s $ and $(M-L)\cap L_{b}=\phi $;

(U5) Bob checks $b=c\odot r$ and ($c$ is a codeword).

\subsection{Notes}

Since it is an important theoretical problem whether secure QBC exists,
here the feasibility of the protocol is not what we care of. Thus we do not
consider the presence of detection error, channel noise, or any other
implementation issue.

In Ref. \cite{He} we used to require Bob to choose $0<f_{a},$\ $f_{b},$\ $%
f_{c}<1/4$\ in step (C3). The purpose is to prevent Bob from delaying his
measurement too often, because if he announces the \textquotedblleft
fake\textquotedblright\ result $\left\vert p_{i}^{\prime \prime
},q_{i}^{\prime \prime }\right\rangle _{\beta }$\ ($i\in S$) before he
actually performs the measurement and obtains the real outcome $\left\vert
p_{i}^{\prime },q_{i}^{\prime }\right\rangle _{\beta }$, then there will be
no specific relationship between $\left\vert p_{i}^{\prime \prime
},q_{i}^{\prime \prime }\right\rangle _{\beta }$\ and $\left\vert
p_{i}^{\prime },q_{i}^{\prime }\right\rangle _{\beta }$, which is equivalent
to choosing $f_{a}=f_{b}=f_{c}=1/4$. By the time Ref. \cite{He} was written
we did not know whether Bob will be benefited if he delays the measurement,
so we introduced the requirement $0<f_{a},$\ $f_{b},$\ $f_{c}<1/4$. But now
we know that Bob cannot cheat even if the measurement was delayed, as it
will be elaborated later in this paper. Thus we can remove this requirement
from now on.

\subsection{An easy understanding}

As the no-go proofs has been widely accepted for more than a decade and a
half, if there is a loophole, it must be lying somewhere subtle. Thus it is
not surprising that a counter-example would look very complicated. To fully
understand how the above protocol works, it is strongly recommend to read
Ref. \cite{He} in detail. For easier comprehension, some main ideas will be
outlined below. But for any security debate in the future, it is important
to always get back to the above rigorous mathematical description, as the
security of a protocol will depend heavily on its details.

The main part of the above commit protocol is to force Alice to accomplish a
lie-detecting task. That is, Alice sends Bob $s$ quantum registers $\beta
_{i}$ ($i=1,...,s$) in step (C1). Bob measures them in (C2) and announces
the results in (C3). But it is important to note that the protocol allows
Bob to lie when announcing the results. Then in (C4), Alice is required to
detect Bob's lies and announces the label $i$ whenever she finds that Bob's
announced result for $\beta _{i}$ is a lie. The total number of lies she is
required to detect is%
\begin{equation}
l\equiv \left\vert L\right\vert \simeq (f_{a}/2+f_{b}/4+f_{c}/4)s.
\label{lie}
\end{equation}%
Here $f_{a}$, $f_{b}$, and $f_{c}$ are the lying frequencies with which Bob
announces different types of lies. A type $a$ lie means that Bob announces his
actual measurement basis $p_{i}^{\prime }$ honestly as $%
p_{i}^{\prime \prime }$, while lies about the state he found in this basis.
That is, when his actual measurement result is $\left\vert p_{i}^{\prime
},q_{i}^{\prime }\right\rangle _{\beta }$, he takes $\left\vert
p_{i}^{\prime \prime },q_{i}^{\prime \prime }\right\rangle _{\beta }=
\left\vert p_{i}^{\prime },\urcorner q_{i}^{\prime }\right\rangle _{\beta }$
and announces $\left\vert
p_{i}^{\prime \prime },q_{i}^{\prime \prime }\right\rangle _{\beta }$
. On the contrary, a type $b$ lie means that Bob lies about the basis, while
announcing $q_{i}^{\prime }$ honestly, i.e., the actual result $\left\vert
p_{i}^{\prime },q_{i}^{\prime }\right\rangle _{\beta }$ is announced as $%
\left\vert p_{i}^{\prime \prime },q_{i}^{\prime \prime }\right\rangle
_{\beta }=\left\vert \urcorner p_{i}^{\prime },q_{i}^{\prime
}\right\rangle _{\beta }$ instead.\ A type $c$ lie means that Bob lies about both the
basis $p_{i}^{\prime }$ and the state $q_{i}^{\prime }$, i.e., the actual
result $\left\vert p_{i}^{\prime },q_{i}^{\prime }\right\rangle _{\beta }$
is announced as $\left\vert p_{i}^{\prime \prime },q_{i}^{\prime \prime
}\right\rangle _{\beta }=\left\vert \urcorner p_{i}^{\prime },\urcorner
q_{i}^{\prime }\right\rangle _{\beta }$ instead. Note that if a \textquotedblleft
fake\textquotedblright\ result $\left\vert p_{i}^{\prime \prime
},q_{i}^{\prime \prime }\right\rangle _{\beta }$ Bob announced is not a lie
at all, it will be called an honest result when we need to distinguish it
from other lies. But in general, for simplicity we will still call
everything (either lies or honest ones) denoted by $\left\vert
p_{i}^{\prime \prime },q_{i}^{\prime \prime }\right\rangle _{\beta }$ as
fake results.

The commit protocol not only require Alice to detect $l$ lies, but also
force her to use the optimal strategy. Here \textquotedblleft
optimal\textquotedblright\ means that while the total number of detected
lies must reach $l$, Alice should try her best to keep the number of the
unmeasured quantum registers $\alpha _{i}$ as large as possible, so that
most $\alpha _{i}\otimes \beta _{i}$\ pairs remain entangled \cite{note1},
while only a small portion of them was measured and collapsed into
non-entangled product states. As shown in Ref. \cite{He}, when $%
f_{a}+f_{c}<1/2$ and $f_{b}>f_{c}$,\ the optimal strategy for Alice is to
prepare the initial states of $\alpha _{i}\otimes \beta _{i}$ in a
non-maximally entangled form as Eq. (\ref{eqA1}). Then to detect $l$ lies,
the number of $\alpha _{i}$ she needs to measure is as small as
\begin{equation}
m\equiv \left\vert M\right\vert \simeq \lbrack 1/4+(f_{a}+f_{c})/2]s.
\label{measured}
\end{equation}%
Therefore, after $l$ lies were detected and the corresponding quantum
registers were discarded, the remaining $n=s-l$ pairs of quantum registers
contain $m-l$ pairs of measured ones, while the rest $n-(m-l)=s-m$ pairs
remain entangled from Alice's point of view as she has not measured the
corresponding $\alpha _{i}$. By assigning a \textquotedblleft $0$%
\textquotedblright\ to each of the unmeasured ones and \textquotedblleft $1$%
\textquotedblright\ to each of the measured ones, respectively, Alice obtains
an $n$-bit string $c^{0}$ in step (C6). As it is a basic law that
entanglement cannot be created locally, Alice cannot change the
\textquotedblleft $1$\textquotedblright\ to \textquotedblleft $0$%
\textquotedblright\ in $c^{0}$ freely. Step (C7) further connects $c^{0}$
with Alice's commit bit $b$. Thus Alice is forced to commit once she
accomplishes the lie-detecting task.

\section{The density matrix in our protocol}

\subsection{Important hints}

When calculating the density matrix $\rho _{b}^{B}$, two things should be
kept in mind.

(i) We only need to study the value when the participants act honestly. This
may look weird at the first glance. But we should note that the conclusion
of the no-go proofs is: if $\rho _{0}^{B}\simeq \rho _{1}^{B}$\ is satisfied
when both participants execute the protocol honestly, then Alice can cheat.
That is, the density matrix $\rho _{b}^{B}$\ studied in the no-go proofs is
the one that describes the state obtained in the honest protocol, before
taking cheating into consideration. In fact, even if Alice cheats, $\rho
_{b}^{B}$\ should remain unchanged. Otherwise Bob can simply perform a
measurement to distinguish the density matrices, thus reveal Alice's
cheating. On the other hand, suppose that Bob cheats by introducing
ancillary systems and/or performing transformations to alter $\rho _{b}^{B}$%
. Then we can always treat all these ancillary systems and transformations
as a part of his operations on distinguishing $\rho _{b}^{B}$, instead of a
part of $\rho _{b}^{B}$\ itself. Therefore, no cheating of either
participant need to be considered when calculating $\rho _{b}^{B}$.

(ii) $\rho _{b}^{B}$ should not only describe the quantum system Alice sent
to Bob (e.g., the registers $\beta _{i}$'s in our protocol), but also
reflect the influence of classical communication. The latter includes the
classical information Alice announces to Bob, as well as what Bob announces
to Alice while she accepts without questioning (i.e., Bob can assume by
default that his classical information has reached Alice successfully so
that she knows the content). This is because the original MLC no-go theorem
worked on a scenario without involving classical communication directly. But
it is by no means indicating that classical communication can be
simply ignored. Instead, they used an \textquotedblleft
indirect\textquotedblright\ approach (as named in Ref. \cite{qi82}). That
is, they treated classical communication as a special case of quantum
communication, and replaced them with a quantum channel \cite{qi105}.
Consequently, any protocol using classical information are replaced with a
full quantum protocol without classical information. As pinpointed out in
section 2 of Ref. \cite{qi82}, the advantage is that the attack on the new
protocol is easy to describe, while the disadvantage is that the attack
obtained against the new protocol is not the one that applies on the
original protocol. Therefore, to make our presentation consistent with the above
description of our QBC protocol (which includes classical communication) so
that it could be easier for the reader to understand, here we avoid using
the indirect approach, and calculate $\rho _{b}^{B}$ with classical
communication taken into account in its original form.

\subsection{The constraint from $\left\vert p_{i},q_{i}\right\rangle _{%
\protect\beta }$}

With the above considerations, let us study the quantum states at the end of
our commit protocol. The informations corresponding to the quantum registers
$\beta _{i}$'s ($i\in L$) were already detected as lies in step (C4) and
were publicly known to both participants, so that they are no longer useful
and can be discarded. Thus we are interested in the remaining $\beta _{i}$'s
($i\in S-L$) at Bob's side. To each of them, Alice has assigned a bit $%
c_{i}^{0}$ in step (C6). Since Alice has not announced Bob's corresponding
fake\ result $\left\vert p_{i}^{\prime \prime },q_{i}^{\prime \prime
}\right\rangle _{\beta }$\ as a lie, it indicates two possibilities
by default.

(a) $c_{i}^{0}=1$, i.e., Alice has measured the corresponding $\alpha _{i}$
in step (C4) but detected no lie.

(b) $c_{i}^{0}=0$, i.e., Alice has chosen not to measure $\alpha _{i}$ in
step (C4).

In step (C4) Alice measures all $\alpha _{i}$'s that satisfy $q_{i}^{\prime
\prime }=\urcorner q_{i}$, and announces these satisfying $%
p_{i}=p_{i}^{\prime \prime }$\ as lies. Therefore according to Eq. (\ref%
{eqA1}), in case (a) Alice's measurement will collapse $\left\vert \psi
_{i}\right\rangle =\left\vert \alpha _{i}\otimes \beta _{i}\right\rangle $
into%
\begin{equation}
\left\vert \psi _{i}\right\rangle \rightarrow \left\{
\begin{array}{c}
\left\vert x\right\rangle _{\alpha }\otimes \left\vert \urcorner
p_{i}^{\prime \prime },\urcorner q_{i}^{\prime \prime }\right\rangle _{\beta
},\text{ \ \ }(p_{i}^{\prime \prime }=1) \\
\left\vert y\right\rangle _{\alpha }\otimes \left\vert \urcorner
p_{i}^{\prime \prime },\urcorner q_{i}^{\prime \prime }\right\rangle _{\beta
},\text{ \ \ }(p_{i}^{\prime \prime }=0)%
\end{array}%
\right.   \label{case a}
\end{equation}%
In case (b), $\left\vert \psi _{i}\right\rangle $ can be written as%
\begin{equation}
\left\vert \psi _{i}\right\rangle =\cos \theta _{i}\left\vert x\right\rangle
_{\alpha }\otimes \left\vert 0,q_{i}^{\prime \prime }\right\rangle _{\beta
}+\sin \theta _{i}\left\vert y\right\rangle _{\alpha }\otimes \left\vert
1,q_{i}^{\prime \prime }\right\rangle _{\beta }.  \label{case b}
\end{equation}%
But these are merely the forms of the states under the constraint of the
relationship between the values of $\left\vert p_{i}^{\prime \prime
},q_{i}^{\prime \prime }\right\rangle _{\beta }$ and $\left\vert
p_{i},q_{i}\right\rangle _{\beta }$. We must further consider the
constraints brought by
the relationship between $\left\vert p_{i}^{\prime
\prime },q_{i}^{\prime \prime }\right\rangle _{\beta }$ and Bob's actual
measurement result.

\subsection{Type $a$ lies}

Suppose that Bob's fake\ result $\left\vert p_{i}^{\prime \prime
},q_{i}^{\prime \prime }\right\rangle _{\beta }$ turns out to be a type $a$
lie, i.e., Bob's actual result is $\left\vert p_{i}^{\prime },q_{i}^{\prime
}\right\rangle _{\beta }=\left\vert p_{i}^{\prime \prime },\urcorner
q_{i}^{\prime \prime }\right\rangle _{\beta }$. Then we can see that from
Alice's point of view, in case (b) among the two components in the
superposition in Eq. (\ref{case b}), the one corresponding to $\left\vert
p_{i}^{\prime \prime },q_{i}^{\prime \prime }\right\rangle _{\beta }$\ will
conflict with Bob's actual result as they are orthogonal. Therefore, though
$c_{i}^{0}=0$\ means
that Alice should keep the entangled state Eq. (\ref{case b}) unmeasured,
this component must vanish when Bob's measurement makes the state collapse.
So the only component that takes effect should be $\left\vert x\right\rangle
_{\alpha }\otimes \left\vert \urcorner p_{i}^{\prime \prime },q_{i}^{\prime
\prime }\right\rangle _{\beta }$ (if $p_{i}^{\prime \prime }=1$) or $%
\left\vert y\right\rangle _{\alpha }\otimes \left\vert \urcorner
p_{i}^{\prime \prime },q_{i}^{\prime \prime }\right\rangle _{\beta }$\ (if $%
p_{i}^{\prime \prime }=0$). That is, if Alice wants to take $c_{i}^{0}=0$,\
then from her point of view (as she does not know Bob's actual result $%
\left\vert p_{i}^{\prime },q_{i}^{\prime }\right\rangle _{\beta }$%
), the state of the corresponding $\beta _{i}$ she sent to Bob has to take
the form $\left\vert \urcorner p_{i}^{\prime \prime },q_{i}^{\prime \prime
}\right\rangle _{\beta }$ at the end of the commit phase. On the other hand, as we
showed above, in case (a) (i.e., if Alice wants to take $c_{i}^{0}=1$) the
state of $\beta _{i}$ should be $\left\vert \urcorner p_{i}^{\prime \prime
},\urcorner q_{i}^{\prime \prime }\right\rangle _{\beta }$. In brief, when
there is a type $a$ lie, the two states of $\beta _{i}$ corresponding to $%
c_{i}^{0}=0$\ and $c_{i}^{0}=1$, respectively, are orthogonal to each other.

Some might wonder why the state at Bob's side is not simply Bob's actual
result $\left\vert p_{i}^{\prime },q_{i}^{\prime }\right\rangle _{\beta }$
itself. This is because, as we mentioned in the above point (ii), the
classical information exchanged in the protocol should also be taken into
consideration. That is, the state at Bob's side that Alice can unveil
successfully later must show no conflict not only with Bob's actual result
$\left\vert p_{i}^{\prime },q_{i}^{\prime }\right\rangle _{\beta }$,
but also with the type of lies that Bob's announced fake result $\left\vert
p_{i}^{\prime \prime },q_{i}^{\prime \prime }\right\rangle _{\beta }$
belongs to, i.e., it should explain why Alice has not detected this lie.
If in the unveil phase Alice said that the state she sent was
$\left\vert p_{i},q_{i}\right\rangle _{\beta }=\left\vert p_{i}^{\prime
\prime },\urcorner q_{i}^{\prime \prime }\right\rangle _{\beta }$, then in
the case of type $a$ lies, it equals exactly to Bob's actual result $%
\left\vert p_{i}^{\prime },q_{i}^{\prime }\right\rangle _{\beta }$. But it
will conflict with $\left\vert p_{i}^{\prime \prime },\urcorner
q_{i}^{\prime \prime }\right\rangle _{\beta }$, because if there is $%
q_{i}=\urcorner q_{i}^{\prime \prime }$, in step (C4) Alice should have
categorized it\ into the measured set $M$. Then as there is also $%
p_{i}=p_{i}^{\prime \prime }$, she should have announced it as a detected
lie in set $L$, which should be discarded in later steps without being
assigned a $c_{i}^{0}$ value at all. Thus we see that Bob's actual result $%
\left\vert p_{i}^{\prime },q_{i}^{\prime }\right\rangle _{\beta }$ cannot be
taken as the state encoding Alice's committed codeword. Instead, as shown in
the previous paragraph, the state $\left\vert \urcorner p_{i}^{\prime \prime
},q_{i}^{\prime \prime }\right\rangle _{\beta }$ (or $\left\vert \urcorner
p_{i}^{\prime \prime },\urcorner q_{i}^{\prime \prime }\right\rangle _{\beta
}$) will not conflict with Bob's actual result $\left\vert p_{i}^{\prime
},q_{i}^{\prime }\right\rangle _{\beta }$, because they are nonorthogonal so
that Bob's measurement can indeed find the result $\left\vert p_{i}^{\prime
},q_{i}^{\prime }\right\rangle _{\beta }$\ with a nonvanishing probability.
Meanwhile, it also agrees with the fact that in step (C4) Alice chose not to
measure the state (or she measured but did not detect it as a lie).
Therefore, it is the correct state at Bob's side at the end of the commit
phase that corresponds to $c_{i}^{0}=0$\ (or $c_{i}^{0}=1$).

\subsection{Type $b$ lies}

Now suppose that Bob's fake\ result $\left\vert p_{i}^{\prime \prime
},q_{i}^{\prime \prime }\right\rangle _{\beta }$ is a type $b$ lie, i.e.,
Bob's actual result is $\left\vert p_{i}^{\prime },q_{i}^{\prime
}\right\rangle _{\beta }=\left\vert \urcorner p_{i}^{\prime \prime
},q_{i}^{\prime \prime }\right\rangle _{\beta }$. Since in step (C4) Alice
only measures the $\alpha _{i}$'s that satisfy $q_{i}^{\prime \prime
}=\urcorner q_{i}$, for a measured $\alpha _{i}$ there will be $%
q_{i}^{\prime \prime }=q_{i}^{\prime }=\urcorner q_{i}$. Rewriting Eq. (\ref%
{eqA1}) as
\begin{eqnarray}
\left\vert \psi _{i}\right\rangle  &=&[\cos \theta _{i}\left\vert
x\right\rangle _{\alpha }+(-1)^{q_{i}}\frac{\sin \theta _{i}}{\sqrt{2}}%
\left\vert y\right\rangle _{\alpha }]\otimes \left\vert 0,q_{i}\right\rangle
_{\beta }  \notag \\
&&+\frac{\sin \theta _{i}}{\sqrt{2}}\left\vert y\right\rangle _{\alpha
}\otimes \left\vert 0,\urcorner q_{i}\right\rangle _{\beta }  \notag \\
&=&\frac{\cos \theta _{i}}{\sqrt{2}}\left\vert x\right\rangle _{\alpha
}\otimes \left\vert 1,\urcorner q_{i}\right\rangle _{\beta }  \notag \\
&&+[(-1)^{q_{i}}\frac{\cos \theta _{i}}{\sqrt{2}}\left\vert x\right\rangle
_{\alpha }+\sin \theta _{i}\left\vert y\right\rangle _{\alpha }]\otimes
\left\vert 1,q_{i}\right\rangle _{\beta }.
\end{eqnarray}%
we can see that in the current case, Bob's measurement in the $p_{i}^{\prime }$ basis
collapses $\left\vert \psi _{i}\right\rangle $ into either $\left\vert
x\right\rangle _{\alpha }\otimes \left\vert 1,\urcorner q_{i}\right\rangle
_{\beta }$ (if $p_{i}^{\prime }=\urcorner p_{i}^{\prime \prime }=1$) or $%
\left\vert y\right\rangle _{\alpha }\otimes \left\vert 0,\urcorner
q_{i}\right\rangle _{\beta }$\ (if $p_{i}^{\prime }=\urcorner p_{i}^{\prime
\prime }=0$). When Alice measures this $\alpha _{i}$ in step (C4), her
result will always be $p_{i}=p_{i}^{\prime \prime }$ so that she would
detect it as a lie. That is, all the type $b$ lies in set $M$ will be
detected. After Alice announced set $L$ in step (C4), there will be no more
type $b$ lie left in the set $M-L$. (This is why Bob needs to check $(M-L)\cap
L_{b}=\phi $ in step (U4).) Therefore at the end of the commit phase, for
any $\beta _{i}$ corresponding to a bit in the string $%
c^{0}=(c_{1}^{0}c_{2}^{0}...c_{n}^{0})$, Bob's fake result must not be a
type $b$ lie if $c_{i}^{0}=1$. Any type $b$ lie has to indicate $c_{i}^{0}=0$%
. In other words, for a specific $i$ if Bob's fake result is a type $b$ lie,
then the $\beta _{i}$ at Bob's side at this stage has to be in the state
corresponding to $c_{i}^{0}=0$. There does not exist any legitimate state of
$\beta _{i}$ that can be later unveiled as $c_{i}^{0}=1$. In this sense, the
states of $\beta _{i}$ corresponding to $c_{i}^{0}=0$\ and $c_{i}^{0}=1$,
respectively, are also orthogonal to each other when there is a type $b$ lie.

The state of $\beta _{i}$ corresponding to $c_{i}^{0}=0$\ in this case can
be a mixture of $\left\vert p_{i}^{\prime \prime },q_{i}^{\prime \prime
}\right\rangle _{\beta }$\ and $\left\vert \urcorner p_{i}^{\prime \prime
},q_{i}^{\prime \prime }\right\rangle _{\beta }$, as none of these
components in Eq. (\ref{case b}) conflicts with Bob's actual result $%
\left\vert p_{i}^{\prime },q_{i}^{\prime }\right\rangle _{\beta }=\left\vert
\urcorner p_{i}^{\prime \prime },q_{i}^{\prime \prime }\right\rangle _{\beta
}$.

\subsection{Type $c$ lies}

The above results on types $a$ and $b$ lies are already sufficient for our
discussion on $\rho _{b}^{B}$ in this section. But for completeness, we
further study the type $c$ lie, i.e., Bob's actual result is $\left\vert
p_{i}^{\prime },q_{i}^{\prime }\right\rangle _{\beta }=\left\vert \urcorner
p_{i}^{\prime \prime },\urcorner q_{i}^{\prime \prime }\right\rangle _{\beta
}$. Then from Alice's point of view, in the above case (b) among the two
components in the superposition in Eq. (\ref{case b}), the one corresponding
to $\left\vert \urcorner p_{i}^{\prime \prime },q_{i}^{\prime \prime
}\right\rangle _{\beta }$\ will conflict with Bob's actual result. Therefore
the only component that takes effect should be $\left\vert x\right\rangle
_{\alpha }\otimes \left\vert p_{i}^{\prime \prime },q_{i}^{\prime \prime
}\right\rangle _{\beta }$ (if $p_{i}^{\prime \prime }=0$) or $\left\vert
y\right\rangle _{\alpha }\otimes \left\vert p_{i}^{\prime \prime
},q_{i}^{\prime \prime }\right\rangle _{\beta }$\ (if $p_{i}^{\prime \prime
}=1$). That is, if Alice wants to take $c_{i}^{0}=0$,\ then the state of $%
\beta _{i}$ should take the form $\left\vert p_{i}^{\prime \prime
},q_{i}^{\prime \prime }\right\rangle _{\beta }$ at the end of the commit
phase. On the other hand, as showed above in case (a), if Alice wants to
take $c_{i}^{0}=1$, the state of $\beta _{i}$ should be $\left\vert
\urcorner p_{i}^{\prime \prime },\urcorner q_{i}^{\prime \prime
}\right\rangle _{\beta }$. Thus we see that, unlike types $a$ and $b$ lies,
when there is a type $c$ lie, the two states of $\beta _{i}$ corresponding
to $c_{i}^{0}=0$\ and $c_{i}^{0}=1$, respectively, are \textit{nonorthogonal}
to each other.

\subsection{Honest results}

Similarly, it can be shown that if Bob announced $\left\vert p_{i}^{\prime
\prime },q_{i}^{\prime \prime }\right\rangle _{\beta }=\left\vert
p_{i}^{\prime },q_{i}^{\prime }\right\rangle _{\beta }$ as an honest result
without lying, then the state of $\beta _{i}$ corresponding to $c_{i}^{0}=0$%
\ can be a mixture of $\left\vert p_{i}^{\prime \prime },q_{i}^{\prime
\prime }\right\rangle _{\beta }$\ and $\left\vert \urcorner p_{i}^{\prime
\prime },q_{i}^{\prime \prime }\right\rangle _{\beta }$, like that of the
type $b$ lies. Meanwhile, there exists a legitimate state corresponding to $%
c_{i}^{0}=1$, which is $\left\vert \urcorner p_{i}^{\prime \prime
},\urcorner q_{i}^{\prime \prime }\right\rangle _{\beta }$, like those of
the types $a$ and $c$ lies. Again, the two states are nonorthogonal.

For explicitness, we briefly summarized the above results in Table I.

%
%
%
%

\begin{table}
\caption{\label{tab:table1}The state of $\beta _{i}$ corresponding to different values of $%
c_{i}^{0}$\ and the type of lies that the fake result $\left\vert
p_{i}^{\prime \prime },q_{i}^{\prime \prime }\right\rangle _{\beta }$
belongs to. See section IV for details.}
\begin{ruledtabular}
\begin{tabular}{ccc}
$\left\vert p_{i}^{\prime \prime },q_{i}^{\prime \prime }\right\rangle
_{\beta }$ & $\beta _{i}(c_{i}^{0}=0)$\ & $\beta _{i}(c_{i}^{0}=1)$\\
\hline

Type $a$ lie & $\left\vert \urcorner p_{i}^{\prime \prime },q_{i}^{\prime
\prime }\right\rangle _{\beta }$\ & $\left\vert \urcorner p_{i}^{\prime
\prime },\urcorner q_{i}^{\prime \prime }\right\rangle _{\beta }$\\

\\

Type $b$ lie & Mixture of & Not available\\

& $\left\vert p_{i}^{\prime \prime },q_{i}^{\prime
\prime }\right\rangle _{\beta }$\ and $\left\vert \urcorner p_{i}^{\prime
\prime },q_{i}^{\prime \prime }\right\rangle _{\beta }$\ & \\

\\

Type $c$ lie & $\left\vert p_{i}^{\prime \prime },q_{i}^{\prime \prime
}\right\rangle _{\beta }$\ & $\left\vert \urcorner p_{i}^{\prime \prime
},\urcorner q_{i}^{\prime \prime }\right\rangle _{\beta }$\\

\\

Honest result & Mixture of & $\left\vert
\urcorner p_{i}^{\prime \prime },\urcorner q_{i}^{\prime \prime
}\right\rangle _{\beta }$\\

& $\left\vert p_{i}^{\prime \prime },q_{i}^{\prime
\prime }\right\rangle _{\beta }$\ and $\left\vert \urcorner p_{i}^{\prime
\prime },q_{i}^{\prime \prime }\right\rangle _{\beta }$\ & \\

\end{tabular}
\end{ruledtabular}
\end{table}

\subsection{Bob's required $d$}

The above discussion is about the state of a single $\beta _{i}$. Now let us
turn to the entire system $B\equiv \bigotimes\limits_{i}\beta _{i}$ ($i\in
S-L$) corresponding to the density matrix $\rho _{b}^{B}$. Note that all the
$\beta _{i}$'s\ in $B$ are not completely independent from each other.
Together they should represent a codeword. In brief, the binary linear $%
(n,k,d)$-code $C$ is a set of classical $n$-bit strings. It contains about
$2^{k}$\ strings in total. {}Each string is called a codeword, carefully
selected from all the $2^{n}$ possible choices of $n$-bit strings, so that
the distance (i.e., the number of different bits) between any two codewords
is not less than $d$. Let $\left\vert B(c)\right\rangle \equiv
\bigotimes\limits_{i}\left\vert \beta _{i}(c_{i}^{0})\right\rangle $ denote
the state of system $B$ at the end of the commit phase that corresponds to a
specific codeword $c$, i.e., the relationship $c^{\prime }=c\oplus c^{0}$ is
satisfied, where $c^{0}=(c_{1}^{0}c_{2}^{0}...c_{n}^{0})$ is the string that
the state $\left\vert B(c)\right\rangle $ represents according to the coding
method in step (C6), and $c^{\prime }$\ is what Alice announces in step
(C7.4). Let $\left\vert B(c^{\ast })\right\rangle \equiv
\bigotimes\limits_{i}\left\vert \beta _{i}(c_{i}^{0\ast })\right\rangle $
denote such a state of system $B$ that corresponds to another codeword $%
c^{\ast }$, with $c^{\prime }=c^{\ast }\oplus c^{0\ast }$. Note that $c$ and
$c^{\ast }$ have at least $d$ different bits. According to the analysis
above, if Bob's fake result on one of these $d$ bits (denote it as the $%
\tilde{\imath}$-th bit) is a type $a$ or $b$ lie, the states $\left\vert
\beta _{\tilde{\imath}}(c_{\tilde{\imath}}^{0})\right\rangle $\ and $%
\left\vert \beta _{\tilde{\imath}}(c_{\tilde{\imath}}^{0\ast })\right\rangle
$\ are orthogonal to each other since $c_{\tilde{\imath}}^{0}\neq c_{\tilde{%
\imath}}^{0\ast }$. In this case $\left\langle B(c)\right. \left\vert
B(c^{\ast })\right\rangle =0$ will be rigorously satisfied regardless the
state of other $\beta _{i}$'s ($i\neq \tilde{\imath}$).

It is easy to ensure that the $d$ different bits between any two codewords
contain at least one bit which is corresponding to a type $a$ or $b$ lie.
According to step (C3) of our protocol, the numbers of the types $a$, $b$
and $c$ lies are about $f_{a}s$, $f_{b}s$ and $f_{c}s$, respectively. Eq. (%
\ref{lie}) shows that the numbers of each type of the lies that Alice
detected in step (C4)\ are about $f_{a}s/2$, $f_{b}s/4$ and $f_{c}s/4$,
respectively. Therefore after the commit phase, the numbers of these three
types of lies that remain undetected are about%
\begin{equation}
l_{a}^{\prime }\simeq f_{a}s/2,l_{b}^{\prime }\simeq 3f_{b}s/4,l_{c}^{\prime
}\simeq 3f_{c}s/4.  \label{remained lies}
\end{equation}%
Meanwhile, the total number of honest results are about%
\begin{equation}
h\simeq (1-f_{a}-f_{b}-f_{c})s.
\end{equation}%
Note that the above numbers are all evaluated statistically. Some
fluctuation around these statistical values must be allowed. But the
tolerable range of fluctuation (that can ensure the protocol work properly
with a very high probability) can be estimated using classical statistical
theory, so we are not going into detail here. Obviously, the above numbers
satisfy the following relationship regardless of statistical fluctuation%
\begin{equation}
l_{a}^{\prime }+l_{b}^{\prime }+l_{c}^{\prime }+h=n.
\end{equation}%
Given that Bob's choice of the type of lies is fixed for each and every $%
\beta _{i}$ ($i\in S-L$), then if $d$ is larger than the total numbers of
honest results (i.e., $h$) and the type $c$ lies left undetected (i.e., $%
l_{c}^{\prime }$), and the difference is significantly larger than the
tolerable range of statistical fluctuation, we can be sure that among the $d$
different bits between any two codewords, there is at least one bit that
corresponding to a type $a$ or $b$ lie. Therefore, in step (C7.1) Bob tends
to accept a value of $d$ that satisfies%
\begin{equation}
d>d_{\min }\equiv h+l_{c}^{\prime }\simeq (1-f_{a}-f_{b}-f_{c}/4)s.
\label{Bob d}
\end{equation}%
With this $d$, $\left\langle B(c)\right. \left\vert B(c^{\ast
})\right\rangle =0$ will always be satisfied for any two codewords $c$ and $%
c^{\ast }$. Thus the two Hilbert spaces $H_{0}$ and $H_{1}$\ are rigorously
orthogonal to each other, where $H_{b}$\ ($b=0,1$) denotes the space
supported by all the states $\left\vert B(c)\right\rangle $'s that
corresponding to those codewords $c$'s which can unveil the value of the
committed bit as $b$ successfully, i.e., $\{c\in C|c\odot r=b\}$. Using $%
\lambda _{c}$ to denote the probability for a codeword $c$ to be chosen when
Alice's committed value is $b$\ ($b=0,1$), we have $\rho
_{b}^{B}=\sum\nolimits_{c\in C|c\odot r=b}\lambda _{c}\left\vert
B(c)\right\rangle \left\langle B(c)\right\vert $. Then we reach one of the
main conclusion of the current paper, that there will be $\rho _{0}^{B}\perp
\rho _{1}^{B}$\ as long as $d$ satisfies Eq. (\ref{Bob d}). In this case,
Alice cannot cheat using the specific strategy proposed in the existing
no-go theorem of QBC, because it has to rely on the condition $\rho
_{0}^{B}\simeq \rho _{1}^{B}$.

\section{Security against Bob's cheating}

\subsection{Bob's dilemma: which basis to measure}

It remains to show that our protocol is still concealing even though $\rho
_{0}^{B}\perp \rho _{1}^{B}$. At the first glance it seems impossible,
because Bob can simple perform a collective measurement that distinguishes $%
\rho _{0}^{B}$ from $\rho _{1}^{B}$\ and learn the committed bit $b$. More
rigorously, Bob finds out all the codewords $c\in C$ that satisfy $c\odot r=0
$, then constructs the projection operator%
\begin{equation}
P_{0}\equiv \sum\nolimits_{c\in C|c\odot r=0}\left\vert B(c)\right\rangle
\left\langle B(c)\right\vert ,  \label{projector}
\end{equation}%
and applies it on his quantum system $B=\bigotimes\limits_{i}\beta _{i}$ ($%
i\in S-L$) after the commit phase. If the projection is successful, he knows
that $b=0$. Else if the projection fails, he knows that $b=1$.

However, we must note that $\rho _{0}^{B}\perp \rho _{1}^{B}$\ is
conditional. It requires Eq. (\ref{Bob d}), which is derived under the
assumption that Bob's choice of the type of lies for every $\beta _{i}$ ($%
i\in S-L$) is already fixed. To calculate the operator $P_{0}$ in Eq. (\ref%
{projector}), Bob must know the form of the state $\left\vert
B(c)\right\rangle =\bigotimes\limits_{i}\left\vert \beta
_{i}(c_{i}^{0})\right\rangle $ corresponding to each codewords $c$
satisfying $c\odot r=0$. According to Table I, $\left\vert \beta
_{i}(c_{i}^{0})\right\rangle $ has different forms for different types of
lies. Without knowing the choice of the type of lies for each $\beta _{i}$, $%
P_{0}$ cannot be obtained.

On the other hand, suppose that Bob tries to calculate $P_{0}$ without
fixing the type of lies. That is, he exhausts all possible combinations of
the types of lies, finds the form of $\left\vert B(c)\right\rangle $
corresponding to each of these combinations, and includes all these $%
\left\vert B(c)\right\rangle $'s into the sum in Eq. (\ref{projector}). Then
the resultant $P_{0}$ will be useless for the reason below. As the choice of
the type of lies is not fixed, suppose that we first calculate the form of $%
\left\vert B(c)\right\rangle $\ by assuming that the fake result $\left\vert
p_{i}^{\prime \prime },q_{i}^{\prime \prime }\right\rangle _{\beta }$\ for $%
\beta _{i}$ is a type $a$ lie when $i=1$, and it is a type $c$ lie when $i=2$%
, ... , then we calculate the form of $\left\vert B(c^{\ast })\right\rangle $%
\ corresponding to a different codeword $c^{\ast }$\ by assuming that the
fake result\ for $\beta _{i}$ is a type $b$ lie when $i=1$, and it is an
honest result when $i=2$, ... In this case, for any given $i$, we cannot
guarantee that $c_{i}$ and $c_{i}^{\ast }$ are corresponding to the same
type of lies. So we can no longer make the assertion above Eq. (\ref{Bob d}%
), that \textquotedblleft among the $d$ different bits between any two
codewords, there is at least one bit that corresponding to a type $a$ or $b$
lie\textquotedblright , even if we take $d>h+l_{c}^{\prime }$. Consequently,
$\left\langle B(c)\right. \left\vert B(c^{\ast })\right\rangle =0$ will not
necessarily hold, especially when we exhausts all possible combinations of
the types of lies. In turns, $\rho _{0}^{B}\perp \rho _{1}^{B}$\ will become
invalid. Also, there will be some codewords that lead to $\left\langle
B(c)\right. \left\vert B(c^{\ast })\right\rangle \neq 0$, even if $c\odot
r=0 $\ while $c^{\ast }\odot r=1$. Then although Eq. (\ref{projector}) sums
over all $c$ satisfying $c\odot r=0$\ only, the operator $P_{0}$ thus
obtained will actually contains components like $\left\vert B(c^{\ast
})\right\rangle \left\langle B(c^{\ast })\right\vert $\ where $c^{\ast }\odot r=1$.
Applying such a $P_{0}$ will no longer provide the value of $b$, no matter
the projection is successful or not.

Thus we show that to construct the projection operator for the measurement
to distinguish $\rho _{b}^{B}$, Bob needs to know the type of lies
corresponding to every $\beta _{i}$ first. But as we know, the types of lies
are defined according to the comparison between Bob's announced $\left\vert
p_{i}^{\prime \prime },q_{i}^{\prime \prime }\right\rangle _{\beta }$ and
his actual result $\left\vert p_{i}^{\prime },q_{i}^{\prime }\right\rangle
_{\beta }$. Consequently, it brings an important feature to our protocol,
that even Bob himself does not know the type of lies corresponding to each $%
i $ if he has not performed any measurement to obtain $\left\vert
p_{i}^{\prime },q_{i}^{\prime }\right\rangle _{\beta }$ yet. To know this
information, Bob is forced to measure the quantum system at his side. But
when he did, the state is collapsed in the measurement, so that it is
impossible to further perform measurement in another basis to obtain
additional information.

Let us elaborate this in more details. Remind that the type $a$ lie is
defined as $\left\vert p_{i}^{\prime \prime },q_{i}^{\prime \prime
}\right\rangle _{\beta }=$ $\left\vert p_{i}^{\prime },\urcorner
q_{i}^{\prime }\right\rangle _{\beta }$. To identify whether an announced $%
\left\vert p_{i}^{\prime \prime },q_{i}^{\prime \prime }\right\rangle
_{\beta }$ is a type $a$ lie, Bob has to measure $\beta _{i}$ in the\ basis $%
p_{i}^{\prime }=p_{i}^{\prime \prime }$. But for a type $a$ lie, the two
states of $\beta _{i}$ corresponding to $c_{i}^{0}=0$\ and $c_{i}^{0}=1$,
respectively, are $\left\vert \urcorner p_{i}^{\prime \prime },q_{i}^{\prime
\prime }\right\rangle _{\beta }$\ \& $\left\vert \urcorner p_{i}^{\prime
\prime },\urcorner q_{i}^{\prime \prime }\right\rangle _{\beta }$. To
distinguish the two states, the required measurement should be performed in
the $\urcorner p_{i}^{\prime \prime }$\ basis, which is different from $%
p_{i}^{\prime }$. As the two measurements are not commutative, even if Bob
delays his measurement during the commit phase, later when he wants to
decode the committed bit $b$ without Alice's helping, he will face the
dilemma which basis to use for his measurement. To know how the states of $%
\beta _{i}$ corresponding to $c_{i}^{0}=0$\ and $c_{i}^{0}=1$ are defined
for each $i$, he needs to identify the type of lies first. But once he
measured $\beta _{i}$ in the $p_{i}^{\prime \prime }$ basis and found that
it is corresponding to a type $a$ lie, then he lost the chance to perform
the measurement in the $\urcorner p_{i}^{\prime \prime }$\ basis for
identifying $c_{i}^{0}$. In this case, he can no longer know the form of the
state $\left\vert B(c)\right\rangle $, nor the density matrix $\rho
_{b}^{B}=\sum\nolimits_{c\in C|c\odot r=b}\lambda _{c}\left\vert
B(c)\right\rangle \left\langle B(c)\right\vert $. Consequently, he will not
know how to construct $P_{0}$ in Eq. (\ref{projector}) or any other
measurement to decode Alice's committed $b$. This is also the case for type $%
c$ lies and honest results, because there does not exist a single basis that
can distinguish the type of lies and the value of $c_{i}^{0}$\
simultaneously.

In short, though there is $\rho _{0}^{B}\perp \rho _{1}^{B}$\ when Eq. (\ref%
{Bob d}) is satisfied, constructing the measurement operator for
distinguishing $\rho _{b}^{B}$ requires the knowledge on how $\rho _{b}^{B}$
are defined. But when there are types $a$ and $c$ lies and honest results,
the definition of $\rho _{b}^{B}$ is unknown to Bob unless he performs
another measurement to identify the types of lies. As the two measurements
are not commutative, Bob cannot have the best of both worlds. The only
exception, however, is type $b$ lies, which will be studied below.

%

\subsection{Alice's required $d$}

Unlike other types of lies and honest results, once a type $b$ lie is
identified, the value of $c_{i}^{0}$\ will be known to Bob automatically
without requiring another measurement. As shown in Table I, there is no
legitimate state of $\beta _{i}$ that can be unveiled as $c_{i}^{0}=1$. If
Bob announced a fake result $\left\vert p_{i}^{\prime \prime },q_{i}^{\prime
\prime }\right\rangle _{\beta }$ as a type $b$ lie, i.e. his measurement was
performed in the $\urcorner p_{i}^{\prime \prime }$ basis and the actual
result is $\left\vert p_{i}^{\prime },q_{i}^{\prime }\right\rangle _{\beta
}=\left\vert \urcorner p_{i}^{\prime \prime },q_{i}^{\prime \prime
}\right\rangle _{\beta }$, Bob will know that there must be $c_{i}^{0}=0$
once Alice has not announced this $\left\vert p_{i}^{\prime \prime
},q_{i}^{\prime \prime }\right\rangle _{\beta }$ as a detected lie in step
(C4). No need for a further measurement in a different basis. Therefore Bob
would like to maximize the number of the type $b$ lies, to save himself from
the dilemma brought by types $a$ and $c$ lies and honest results. When the
type $b$ lies occur with a sufficiently high frequency $f_{b}$\ so that
there is $l_{b}^{\prime }>n-d$, then he knows more than $n-d$ bits of the
codeword Alice chose. Here $l_{b}^{\prime }$ is the number of the type $b$
lies remained undetected in the commit phase, as presented in Eq. (\ref%
{remained lies}). Since the distance between any two codewords is not less
than $d$, there will be only one codeword in $C$ which contains these bits
known to Bob. Thus he can deduce the rest unknown bits, and learn the
complete codeword so that the value of Alice's committed bit $b$ can be
deduced.

However, this can be avoided by setting an upperbound for $d$.
Although Alice does not know the respective values of $f_{a}$, $f_{b}$, and $%
f_{c} $ that Bob chose, in step (C4) she knows $\left\vert M\right\vert $,
i.e., Eq. (\ref{measured}). Suppose that in step (C7.1) Alice accepts a
value of $d $ that satisfies%
\begin{eqnarray}
d &<&d_{\max }\equiv m-s/4\simeq (f_{a}+f_{c})s/2  \notag \\
&\leq &f_{a}s/2+3f_{c}s/4=l_{a}^{\prime }+l_{c}^{\prime }.  \label{Alice d}
\end{eqnarray}%
That is, $d$ is smaller than the total of types $a$ and $c$ lies that remain
undetected after the commit phase. In this case, among any $n-d$ bits of the
codeword, there will always be at least one bit that corresponds to a type $%
a $ or $c$ lie, i.e., $l_{b}^{\prime }>n-d$\ will never be satisfied. Then
Bob will not have enough type $b$ lies to deduce the complete
codeword, because with more than $d$ bits remained unknown, the possible
choices for codewords will increase exponentially with the value of $k$ of
the $(n,k,d)$-code $C$. As a result, once Eq. (\ref{Alice d}) is met, the
protocol is concealing no matter how Bob chooses his lying frequencies $%
f_{a} $, $f_{b}$, and $f_{c}$.

\subsection{The existence of $d$}

So we can see that in step (C7.1) of the commit protocol, for their own
benefit, Alice will try to lower the value of $d$ so that Eq. (\ref{Alice d}%
) can be satisfied, while Bob will try to increase $d$ to meet Eq. (\ref{Bob
d}). Luckily we can have the best of both worlds. Since%
\begin{eqnarray}
d_{\max }-d_{\min } &=&(m-s/4)-(h+l_{c}^{\prime })  \notag \\
&=&(3f_{a}/2+f_{b}+3f_{c}/4-1)s,
\end{eqnarray}%
there will be $d_{\max }>d_{\min }$\ when%
\begin{equation}
3f_{a}/2+f_{b}+3f_{c}/4>1,  \label{condition}
\end{equation}%
so that there can be a value of $d$ between $d_{\max }$\ and $d_{\min }$\
that both Alice and Bob are willing to accept. Although the above values are
estimated statistically and subjected to fluctuations, when $s$ is very
large there will be enough gap between $d_{\max }$\ and $d_{\min }$\ for
Alice and Bob to choose a proper $d$.

The condition Eq. (\ref{condition}) can be met easily. For example, a simple
choice is that Bob takes $f_{b}>1-3f_{a}/2$\ and $f_{c}=0$\ (in fact any $%
f_{c}$\ satisfying $0\leq f_{c}<1-f_{a}-f_{b}$ will do) in step (C3). Note
that even if Bob does not choose these values honestly, Alice does not need
to worry. All she need is to insist on choosing a value of $d$ that
satisfies Eq. (\ref{Alice d}) in step (C7.1), which is a legitimate action
for an honest Alice that Bob cannot refuse. In this case she can still be
sure that Bob does not have enough type $b$ lies to deduce the codeword.
Meanwhile, if the dishonest Bob accepts such a value of $d$ in order to avoid Alice
feeling suspicious, then Eq. (\ref{Bob d}) may not be satisfied, In this
case $\rho _{0}^{B}\perp \rho _{1}^{B}$\ will no longer be ensured, which
may leave room for Alice's potential cheating. Thus we see that if Bob does
not choose $f_{a}$, $f_{b}$, and $f_{c}$ following Eq. (\ref{condition})
honestly, then he can only hurt his own benefit.

\section{Summary and remarks}

We show above that our protocol satisfies $\rho _{0}^{B}\perp \rho _{1}^{B}$
when Eq. (\ref{Bob d}) is met. Therefore Alice cannot cheat with the
specific strategy presented in existing no-go proofs of unconditionally
secure QBC. The key reason is that all these proofs are based on the HJW
theorem, which requires $\rho _{0}^{B}\simeq \rho _{1}^{B}$.

On the other hand, our protocol remains secure against Bob's cheating when
Eq. (\ref{Alice d}) is met, because the measurement basis for the
discrimination between $\rho _{0}^{B}$ and $\rho _{1}^{B}$ is different from
the basis for learning the definition of $\rho _{0}^{B}$ and $\rho _{1}^{B}$%
. Thus we see that $\rho _{0}^{B}\simeq \rho _{1}^{B}$ is not a necessary
condition for a QBC protocol to be concealing.

Also, Eq. (\ref{condition}) ensures that Eqs. (\ref{Bob d}) and (\ref{Alice
d}) can be satisfied simultaneously, so that the security against both sides
can be guaranteed.

It is worth noting that $\rho _{0}^{B}\perp \rho _{1}^{B}$ actually can also
be found in some unconditionally secure relativistic bit commitment protocols \cite%
{qi44,qi582}, where the committed values are encoded with classical data
instead of quantum states. As pointed out in the 3rd paragraph of the
introduction of \cite{qbc35}, \textquotedblleft Kent's relativistic bit
commitment protocol does not rely on the existence of alternative
decompositions of a density operator, and so its security is not challenged
by the Mayers-Lo-Chau result.\textquotedblright\ They make use of
relativistic constraints to achieve the security against Bob's cheating.
Another previous QBC proposal of ours \cite{HeJPA} have the feature $\rho
_{0}^{B}\perp \rho _{1}^{B}$ too, as it is built on top of a quantum key
distribution scheme based on orthogonal states \cite{qi858}. Whether
relativity is the key of its security is arguable \cite{qi859,qi860}. Our
current protocol is completely free from the need of relativity. Its security
against Bob is provided by keeping the definition of $\rho _{b}^{B}$ secret
from him at the beginning.

The fact that $\rho _{b}^{B}$ is unknown without measurement is also an
interesting feature that sets our protocol apart from the QBC model studied
in many no-go proofs. In the own words of \cite{qi23} (as stated below its
Eq. (2)), \textquotedblleft both Alice and Bob are supposed to know the
states $\left\vert 0\right\rangle $ and $\left\vert 1\right\rangle $. This
implies, in particular, that both of them know the states $\left\vert \phi
_{i}\right\rangle _{B}$\ and $\left\vert \phi _{j}^{\prime }\right\rangle
_{B}$\textquotedblright . Here $\left\vert 0\right\rangle $ ($\left\vert
1\right\rangle $) and $\left\vert \phi _{i}\right\rangle _{B}$\ ($\left\vert
\phi _{j}^{\prime }\right\rangle _{B}$) have the same meanings as these of $%
\left\vert \psi _{b}\right\rangle $ and $\left\vert f_{j}^{(b)}\right\rangle
_{B}$\ ($b=0,1$) in our Eq. (\ref{eqMLC}). In many other no-go proofs,
though it is not explicitly stated, we can still see from the details of
their proofs that they hold the same viewpoint. But as already pinpointed
out in another no-go proof \cite{qi240,qi283}, previously \textquotedblleft
the no-go theorem asserts (this) without proof\textquotedblright , and
\textquotedblleft this assertion is actually not
meaningful\textquotedblright . (An amendment to this problem was made in
\cite{qi240,qi283}. But it only considered the case where the states are
unknown to Alice, instead of Bob, and the proof is still based on $\rho
_{0}^{B}\simeq \rho _{1}^{B}$.)

These findings reveal that the existing impossibility proofs are not
sufficiently general. If unconditionally secure QBC is indeed impossible,
then it is necessary to show that there exists a more universal cheating
strategy which does not rely on the condition $\rho _{0}^{B}\simeq \rho
_{1}^{B}$.

Finally, it is worth noting that entanglement plays an essential role in our
protocol. In many previous QBC protocols proven insecure by the no-go
theorem, the honest participants can execute the protocol successfully by
exchanging pure states unentangled with any system at their sides.
Entanglement is needed only when cheating. On the contrary, in our protocol
even an honest Alice must make use of entangled states to accomplish the
optimal strategy to detect Bob's lies. If she prepares every $\beta _{i}$\
as a pure state $\left\vert p_{i},q_{i}\right\rangle _{\beta }$\ instead,
and sends it to Bob without entangling it with any system at her side, then
she cannot detect the lies with the efficiency required in the protocol.
Therefore such an Alice will be caught as cheating instead of honest. That
is, our protocol cannot be accomplished without entanglement. Since
entanglement is a typical example of nonlocality, this result is consistent
with the claim that nonlocality is necessary for secure QBC, as shown in
section 7 of \cite{HeJPA}, as well as in \cite{qbc85}.

The work was supported in part by the NSF of China under grant No. 10975198,
the NSF of Guangdong province, and the Foundation of Zhongshan University
Advanced Research Center.


\begin{thebibliography}{99}
\bibitem{qi74} D. Mayers, \textit{quant-ph/9603015v3}. The trouble with
quantum bit commitment


\bibitem{qi24} D. Mayers, \textit{Phys. Rev. Lett.} \textbf{78}, 3414
(1997). Unconditionally secure quantum bit commitment is impossible

\bibitem{qi23} H. -K. Lo and H. F. Chau, \textit{Phys. Rev. Lett.} \textbf{78%
}, 3410 (1997). Is quantum bit commitment really possible?

\bibitem{qi56} C. Cr\'{e}peau, in \textit{Proc. Pragocrypt '96: 1st
International Conference on the Theory and Applications of Cryptology}
(Czech Technical University Publishing House, Prague, 1996). What is going
on with quantum bit commitment?

\bibitem{qi58} H. -K. Lo and H. F. Chau, \textit{Physica D} \textbf{120},
177 (1998). \textit{quant-ph/9605026v2}. Why quantum bit commitment and ideal quantum coin tossing are impossible

\bibitem{qi105} H. F. Chau and H. -K. Lo, \textit{Fortsch. Phys.} \textbf{46}%
, 507 (1998). \textit{quant-ph/9709053v2}. Making an empty promise with a
quantum computer


\bibitem{qi82} G. Brassard, C. Cr\'{e}peau, D. Mayers, and L. Salvail,
\textit{quant-ph/9712023v1}. A brief review on the impossibility of quantum
bit commitment

\bibitem{qi47} G. Brassard, C. Cr\'{e}peau, D. Mayers, and L. Salvail,
\textit{quant-ph/9806031v1}. Defeating classical bit commitments with a
quantum computer

\bibitem{qi611} J. Bub, \textit{Found. Phys.} \textbf{31}, 735 (2001). The
quantum bit commitment theorem

\bibitem{qi147} R. W. Spekkens and T. Rudolph, \textit{Phys. Rev. A} \textbf{%
65}, 012310 (2001). Degrees of concealment and bindingness in quantum bit
commitment protocols

\bibitem{qbc49} R. W. Spekkens, and T. Rudolph, \textit{Quant. Inf. Comput.}
\textbf{2}, 66 (2002). \textit{quant-ph/0107042v2}. Optimization of coherent
attacks in generalizations of the BB84 quantum bit commitment protocol

\bibitem{qi101} G. M. D'Ariano, \textit{quant-ph/0209149v1}. The quantum bit
commitment: a finite open system approach for a complete classification of
protocols

\bibitem{qbc3} G. M. D'Ariano, \textit{quant-ph/0209150v1}. In \textit{Proc.
QCM\&C} (Rinton press, Boston, 2002). Shortened version of \textit{%
quant-ph/0209149}. The quantum bit commitment: a complete classification of
protocols

\bibitem{qbc36} D. Mayers, \textit{quant-ph/0212159v2}. Superselection rules
in quantum cryptography

\bibitem{qbc35} H. Halvorson, \textit{J. Math. Phys.} \textbf{45}, 4920
(2004). \textit{quant-ph/0310001v2}. Remote preparation of arbitrary
ensembles and quantum bit commitment

\bibitem{qi610} A. Kitaev, D. Mayers, and J. Preskill, \textit{Phys. Rev. A}
\textbf{69}, 052326 (2004). Superselection rules and quantum protocols

\bibitem{qi240} C. -Y. Cheung, \textit{quant-ph/0508180v2}. In \textit{Proc.
ERATO Conference on Quantum Information Science 2005} (Tokyo, 2005). Secret
parameters in quantum bit commitment

\bibitem{qi283} C. -Y. Cheung, \textit{Int. J. Mod. Phys. B} \textbf{21}, 4271
(2007). \textit{quant-ph/0601206v1}. Insecurity of
quantum bit commitment with secret parameters

\bibitem{qi323} G. M. D'Ariano, D. Kretschmann, D. Schlingemann, and R. F.
Werner, \textit{Phys. Rev. A} \textbf{76}, 032328 (2007). \textit{%
quant-ph/0605224v2}. Reexamination of quantum bit commitment: The possible
and the impossible.

\bibitem{qi715} G. Chiribella, G. M. D'Ariano, P. Perinotti, D. M.
Schlingemann, and R. F. Werner, \textit{Phys. Lett. A} \textbf{377}, 1076
(2013). \textit{arXiv:0905.3801v1}. A short impossibility proof of quantum
bit commitment

\bibitem{qbc12} G. Chiribella, G. M. D'Ariano,\ and P. Perinotti, \textit{%
Phys. Rev. A} \textbf{81}, 062348 (2010). \textit{arXiv:0908.1583v5}.
Probabilistic theories with purification

\bibitem{qi714} L. Magnin, F. Magniez, A. Leverrier, and N. J. Cerf, \textit{%
Phys. Rev. A} \textbf{81}, 010302(R) (2010). \textit{arXiv:0905.3419v2}.
Strong no-go theorem for Gaussian quantum bit commitment

\bibitem{qbc32} Q. Li, C. -Q. Li, D. -Y. Long, W. H. Chan, and C. -H. Wu,
\textit{Quantum Inf. Process.} \textbf{11,} 519 (2012). \textit{%
arXiv:1101.5684v1}. On the impossibility of non-static quantum bit
commitment between two parties

\bibitem{qbc31} A. Chailloux and I. Kerenidis, in \textit{Proc. 52nd IEEE
Symposium on Foundations of Computer Science} (2011), p. 354. \textit{arXiv:1102.1678v1}.
Optimal bounds for quantum bit commitment

\bibitem{HeJPA} G. P. He, \textit{J. Phys. A: Math. Theor.} \textbf{44},
445305 (2011). Quantum key distribution based on orthogonal states allows
secure quantum bit commitment

\bibitem{qi70} P. W. Shor and J. Preskill, \textit{Phys. Rev. Lett.} \textbf{%
85}, 441 (2000). Simple proof of security of the BB84 quantum key
distribution protocol

\bibitem{He} G. P. He, \textit{Phys. Rev. A} \textbf{74}, 022332 (2006). (It
is an extended version of \cite{He arxiv}, with refrained claim on the
security as we could not come up with a general proof at that time. The
protocol remains the same.) Secure quantum bit commitment against empty
promises

\bibitem{He arxiv} G. P. He, \textit{quant-ph/0303107v2}. Quantum bit
commitment using entangled states

\bibitem{steering} E. Schr\"{o}dinger, \textit{Proc. Cambridge Philos. Soc.}
\textbf{31}, 555 (1935). Discussion of probability relations between
separated systems

\bibitem{steering 2} E. Schr\"{o}dinger, \textit{Proc. Cambridge Philos. Soc.%
} \textbf{32}, 446 (1936). Probability relations between separated systems

\bibitem{qbc8} A. Uhlmann, \textit{Rep. Math. Phys.} \textbf{9}, 273 (1976).
The \textquotedblleft transition probability\textquotedblright\ in the state
space of a *-algebra

\bibitem{qi73} L. P. Hughston, R. Jozsa, and W. K. Wootters, \textit{Phys.
Lett. A} \textbf{183}, 14 (1993). A complete classification of quantum
ensembles having a given density matrix

\bibitem{qi43} G. Brassard, C. Cr\'{e}peau, R. Jozsa, and D. Langlois, in
\textit{Proc. the 34th Annual IEEE Symposium on Foundations of Computer
Science} (IEEE, Los Alamitos, 1993), p. 362. A quantum bit commitment scheme
provably unbreakable by both parties

\bibitem{note1} Note that Alice's and Bob's measurements on the entangled
state $\left\vert \alpha _{i}\otimes \beta _{i}\right\rangle $\ are
commutative, and Bob is even allowed to delay his measurement on some of the
quantum registers in step (C2) of the commit protocol. Therefore, from
Alice's point of view, $\alpha _{i}$ and $\beta _{i}$ can still be regarded
as entangled as long as she has not measured $\alpha _{i}$, no matter Bob
has actually measured $\beta _{i}$ or not.

\bibitem{qi44} A. Kent, \textit{Phys. Rev. Lett.} \textbf{83}, 1447 (1999).
\textit{quant-ph/9810068v4}. Unconditionally secure bit commitment

\bibitem{qi582} A. Kent, \textit{J. Cryptol.} \textbf{18}, 313 (2005).
\textit{quant-ph/9906103v7}. Secure classical bit commitment using fixed
capacity communication channels

\bibitem{qi858} L. Goldenberg and L. Vaidman, \textit{Phys. Rev. Lett.}
\textbf{75,} 1239 (1995). Quantum cryptography based on orthogonal states

\bibitem{qi859} A. Peres, \textit{Phys. Rev. Lett.} \textbf{77,} 3264
(1996). Quantum cryptography with orthogonal states?

\bibitem{qi860} L. Goldenberg and L. Vaidman, \textit{Phys. Rev. Lett.}
\textbf{77,} 3265 (1996). Reply to \textquotedblleft Quantum cryptography
with orthogonal states?\textquotedblright

\bibitem{qbc85} G. Murta, M. T. Cunha, and A. Cabello, \textit{%
arXiv:1307.0156v2}. Quantum nonlocality allows for ever-lasting
unconditionally secure bit commitment
\end{thebibliography}
\end{document}